\documentclass[prl,aps,amsfonts,amssymb,amsmath,floats,floatfix,showpacs,twocolumn,superscriptaddress,unsortedaddress]{revtex4}
\usepackage{graphics}
\usepackage{epsfig,graphicx}

\begin{document}

\title{Magnetic Superstructure and Metal-Insulator Transition in Mn-Substituted Sr$_3$Ru$_2$O$_7$}

\author{M.A. Hossain}
\email{hossain@slac.stanford.edu} \affiliation{Department of Physics {\rm {\&}} Astronomy, University of
British Columbia, Vancouver, British Columbia V6T\,1Z1, Canada} \affiliation{Advanced Light Source, Lawrence
Berkeley National Laboratory, Berkeley, California 94720, USA}
\author{B. Bohnenbuck}
\affiliation{Max-Planck-Institut f\"{u}r Festk\"{o}rperforschung, Heisenbergstra\ss e 1, 70569 Stuttgart,
Germany}
\author{Y.-D. Chuang}
\affiliation{Advanced Light Source, Lawrence Berkeley National Laboratory, Berkeley, California 94720, USA}
\author{A.G. Cruz Gonzalez}
\affiliation{Advanced Light Source, Lawrence Berkeley National Laboratory, Berkeley, California 94720, USA}
\author{I. Zegkinoglou}
\affiliation{Max-Planck-Institut f\"{u}r Festk\"{o}rperforschung, Heisenbergstra\ss e 1, 70569 Stuttgart,
Germany}
\author{M.W. Haverkort}
\affiliation{Max-Planck-Institut f\"{u}r Festk\"{o}rperforschung, Heisenbergstra\ss e 1, 70569 Stuttgart,
Germany}
\author{\\J. Geck}
\affiliation{Department of Physics {\rm {\&}} Astronomy, University of British Columbia, Vancouver, British
Columbia V6T\,1Z1, Canada}
\author{D.G. Hawthorn}
\affiliation{Department of Physics {\rm {\&}} Astronomy, University of British Columbia, Vancouver, British
Columbia V6T\,1Z1, Canada}
\author{H.-H. Wu}
\affiliation{II. Physikalisches Institut, Universit\"{a}t zu K\"{o}ln, Z\"{u}lpicher Stra\ss e 77, 50937
K\"{o}ln, Germany}
\author{C. Sch\"{u}\ss ler-Langeheine}
\affiliation{II. Physikalisches Institut, Universit\"{a}t zu K\"{o}ln, Z\"{u}lpicher Stra\ss e 77, 50937
K\"{o}ln, Germany}
\author{R. Mathieu}
\affiliation{Department of Applied Physics, University of Tokyo, Tokyo 113-8656, Japan}
\author{Y. Tokura}
\affiliation{Department of Applied Physics, University of Tokyo, Tokyo 113-8656, Japan}
\author{S. Satow}
\affiliation{Department of Advanced Materials Science, University of Tokyo, Kashiwa, Chiba 277-8581, Japan}
\author{H. Takagi}
\affiliation{Department of Advanced Materials Science, University of Tokyo, Kashiwa, Chiba 277-8581, Japan}
\author{\\Y. Yoshida}
\affiliation{National Institute of Advanced Industrial Science and Technology  (AIST), Tsukuba, 305-8568,
Japan}
\author{J.D. Denlinger}
\affiliation{Advanced Light Source, Lawrence Berkeley National Laboratory, Berkeley, California 94720, USA}
\author{I.S. Elfimov}
\affiliation{Department of Physics {\rm {\&}} Astronomy, University of British Columbia, Vancouver, British
Columbia V6T\,1Z1, Canada}
\author{Z. Hussain}
\affiliation{Advanced Light Source, Lawrence Berkeley National Laboratory, Berkeley, California 94720, USA}
\author{B. Keimer}
\affiliation{Max-Planck-Institut f\"{u}r Festk\"{o}rperforschung, Heisenbergstra\ss e 1, 70569 Stuttgart,
Germany}
\author{G.A. Sawatzky}
\affiliation{Department of Physics {\rm {\&}} Astronomy, University of British Columbia, Vancouver, British
Columbia V6T\,1Z1, Canada}
\author{A. Damascelli}
\email{damascelli@physics.ubc.ca} \affiliation{Department of Physics {\rm {\&}} Astronomy, University of
British Columbia, Vancouver, British Columbia V6T\,1Z1, Canada}

\begin{abstract}
We present a temperature-dependent resonant elastic soft x-ray scattering (REXS) study of the metal-insulator
transition in Sr$_3$(Ru$_{1-x}$Mn$_x$)$_2$O$_7$, performed at both Ru and Mn $L$-edges. Resonant magnetic
superstructure reflections, which indicate an incipient instability of the parent compound, are detected
below the transition. Based on modelling of the REXS intensity from randomly distributed Mn impurities, we
establish the inhomogeneous nature of the metal-insulator transition, with an effective percolation threshold
corresponding \!to \!an \!anomalously \!low \!$x\!\lesssim\!0.05$ \!Mn \!substitution.
\end{abstract}

\date{Received May 7, 2009}

\pacs{\vspace{-0.2cm}71.30.+h,75.25.+z,74.70.Pq}

%\pacs{71.30.+h,75.25.+z,74.70.Pq}

\maketitle

Electronic and lattice instabilities in strongly correlated electron systems give rise to many fascinating
phenomena, such as for example various types of spin, charge, and orbital ordering. This is a common feature
of many of the 3$d$ transition-metal oxides, with the best-known examples including the stripe instability in
the cuprate superconductors and the magnetic phase separation in manganites. Competing instabilities and
ordering phenomena can also be found in the somewhat less correlated 4$d$ transition-metal oxides, with the
ruthenates being one of the most prominent families. Sr$_3$Ru$_2$O$_7$, which is the subject of this study,
is known as a metal on the verge of ferromagnetism \cite{ikeda} due to the presence of strong ferromagnetic
fluctuations. More recently, magnetic field tuned quantum criticality \cite{grigera} and electronic nematic
fluid behavior \cite{borzi} have been proposed for this compound and associated with a metamagnetic
transition. However, the deeper connection between these effects is still highly debated and its description
will depend on a fuller understanding of the incipient instabilities in Sr$_3$Ru$_2$O$_7$.

Magnetic impurities such as Mn have been introduced in Sr$_3$Ru$_2$O$_7$ in an attempt to stabilize the
magnetic order in the system \cite{mathieu}. It has been shown that due to the interplay between localized
Mn\,3$d$ and delocalized Ru\,4$d$\,-\,O\,2$p$ valence states, Mn impurities display an unusual crystal field
level inversion already at room temperature \cite{hossain_PRL}. Upon lowering the temperature, a
metal-insulator phase transition has been observed for 5\,\% Mn substitution at $T_c\!\simeq\!50$\,K, and at
progressively higher $T_c$ upon increasing the Mn concentration \cite{mathieu}. In this Letter, we
investigate the nature of the low-temperature insulating phase of Sr$_3$(Ru$_{1-x}$Mn$_x$)$_2$O$_7$ and the
role of Mn impurities using resonant elastic soft x-ray scattering (REXS). We will show that dilute Mn
impurities not only induce long-range magnetic order but also provide a unique opportunity to probe the
electronic instabilities in the parent compound and thereby reveal the mechanism behind the metal-insulator
transition itself.

A starting point for our REXS study has already been provided by previous neutron scattering work that
detected $\boldsymbol{q}\!=\!(\frac{1}{4},\frac{1}{4},0)$ and $(\frac{1}{4},\frac{3}{4},0)$ superlattice
peaks appearing below $T_c$ for the 5\,\% Mn system \cite{mathieu}. While this suggests the emergence of an
electronic modulation and in particular of magnetic order, to specify its nature we need to establish if the
modulation vector $\boldsymbol{q}$ changes upon varying the Mn concentration beyond $x\!=\!0.05$. In
addition, to pinpoint the role of Mn impurities it is necessary to use an experimental tool that can probe Ru
and Mn selectively. While neutron scattering experiments are sensitive to the average magnetic order, REXS
can probe the order on Ru and Mn independently.

REXS is a relatively new spectroscopic technique to probe and study long-range charge/spin/orbital order in
an element specific and direct way. REXS measurements were performed at beamlines 8.0.1 at ALS in Berkeley
(Mn $L$-edges) and KMC-1 at BESSY in Berlin (Ru $L$-edges). In both cases we used a two-circle
ultra-high-vacuum diffractometer in horizontal scattering geometry, with the incident photon beam polarized
parallel to the diffraction plane ($\pi$). The scattered signal contained polarization components parallel
($\pi'$) and perpendicular ($\sigma'$) to the diffraction plane. Sr$_3$(Ru$_{1-x}$Mn$_x$)$_2$O$_7$ single
crystals grown by the floating zone technique \cite{mathieu} were cut and polished along the (110) direction.
\begin{figure*}[t!]
\centerline{\epsfig{figure=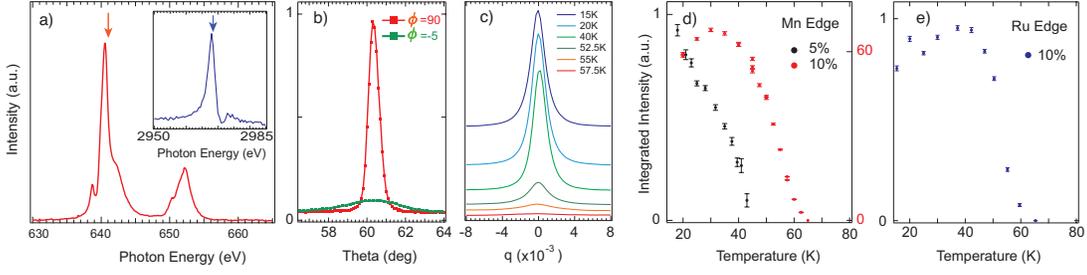,width=.8\linewidth,clip=}} \vspace{-0.25cm}\caption{(color
online). (a) Mn and Ru (inset) resonance profile for the $(\frac{1}{4},\frac{1}{4},0)$ superlattice
diffraction peak measured at 20\,K on Sr$_3$(Ru$_{1-x}$Mn$_x$)$_2$O$_7$ with $x\!=\!0.1$. The arrows at 641
and 2968\,eV indicate the energies used in the REXS experiments. (b) 10\,\% Mn rocking curves for two
different azimuthal angles $\phi$ measured at 641\,eV and 20\,K. (c) 10\,\% Mn $(\frac{1}{4}+\Delta
q,\frac{1}{4}+\Delta q,0)$ momentum scans at 641\,eV for different temperatures. (d) Temperature dependence
of the integrated intensity of the Mn-edge $(\frac{1}{4}+\Delta q,\frac{1}{4}+\Delta q,0)$ momentum scans for
$x\!=\!5$ and 10\,\% samples. (e) Same as (d), for the $x\!=\!10$\,\% sample only, at the Ru
edge.}\label{Fig1}
\end{figure*}
The samples were mounted on cryogenic manipulators, allowing a cryostat polar angle rotation ($\theta$) and
also azimuthal rotation ($\phi$) of the sample about the scattering vector, in the temperature range
20-300\,K. Note that the diffraction peaks will be indexed with respect to the undistorted tetragonal
$I$4$/mmm$ unit cell with axes along the RuO bond directions ($a_0\!=\!b_0\!\simeq\!3.9$\,\AA), while the
magnetic and transport anisotropy will be discussed with reference to the 45$^{\circ}$ rotated and distorted
orthorhombic $Bbcb$ unit cell of Sr$_3$Ru$_2$O$_7$ ($a^*,b^*\!\simeq\!5.5$\,\AA) \cite{Shaked}.

\begin{figure}[b]
\centerline{\epsfig{figure=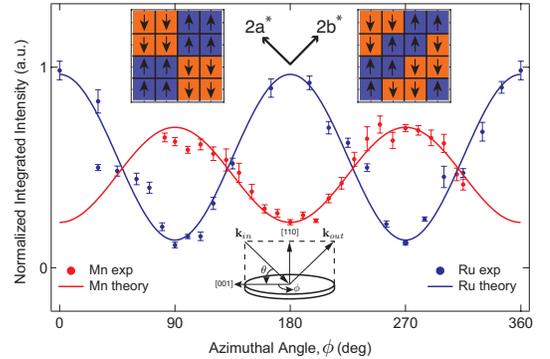,width=.80\linewidth,clip=}} \vspace{-0.25cm}\caption{(color
online). Azimuthal-angle dependence of the $(\frac{1}{4}+\Delta q,\frac{1}{4}+\Delta q,0)$ momentum scan
integrated intensity, at the Ru and Mn edge for the $x\!=\!0.1$ sample. Both datasets are fitted to the
formula $I^{total}_{(\frac{1}{4},\frac{1}{4},0)}\!=\!I^{\pi\rightarrow\sigma'}_{(\frac{1}{4},\frac{1}{4},0)}+
I^{\pi\rightarrow\pi'}_{(\frac{1}{4},\frac{1}{4},0)}\!\propto\!\left|\cos\theta \cos \phi
\right|^2\!+\!\left| \sin 2\theta \sin\phi \right|^2$. Due to the different Mn and Ru scattering angles
($\theta_{Mn}\!=\!61.6^{\circ}$; $\theta_{Ru}\!=\!10.9^{\circ}$), the ratio between
$(\pi\rightarrow\sigma^\prime)$ and $(\pi\rightarrow\pi^\prime)$ scattering signals is also different and the
maximum intensity position is shifted by 90$^{\circ}$ in $\phi$. The experimental geometry and the two
highest symmetry spin patterns consistent with the data are also shown.} \label{Fig2}
\end{figure}

We performed REXS experiments on a range of Mn-substituted Sr$_3$Ru$_2$O$_7$ samples, at the Mn and Ru
$L$-edges (for clarity, mostly data from 10\,\% Mn are shown in Fig.\,\ref{Fig1}). For Mn concentration
$x\!=\!0.05$ and 0.1, this revealed a low-temperature structurally forbidden
$\boldsymbol{q}\!=\!(\frac{1}{4},\frac{1}{4},0)$ superlattice diffraction peak at the Mn $L$-edge. Similarly,
experiments at the Ru $L$-edge on a 10\,\% sample detected $\boldsymbol{q}\!=\!(\frac{1}{4},\frac{1}{4},0)$
and also $\boldsymbol{q}\!=\!(\frac{3}{4},\frac{3}{4},0)$ reflections. Fig.\,\ref{Fig1}(a) and its inset
present the $(\frac{1}{4},\frac{1}{4},0)$ resonance profile at $T\!=\!20$\,K, i.e. the energy dependence of
this superlattice peak intensity at the Mn $L_{2,3}$ and Ru $L_2$-edges. Fig.\,\ref{Fig1}(b) shows the
Mn-edge $(\frac{1}{4},\frac{1}{4},0)$ rocking curves at $T\!=\!20$\,K, i.e. the $\theta$-angle dependence of
the intensity at 641\,eV for different azimuthal angles $\phi$ (see inset of Fig.\,\ref{Fig2} for the
experimental geometry). Since the peak width in momentum is proportional to the inverse of the correlation
length, the sharp $q_{x,y}$ dependence ($\theta$ scan at $\phi\!=\!90^\circ$) and the broad response in $q_z$
($\theta$ scan at $\phi\!=\!-5^\circ$) indicate a two-dimensional order with weak correlation along the
$c$-axis. Lastly, the structurally forbidden reflections are detected only below the DC transport
metal-insulator $T_c$, with a progressively increasing strength upon reducing temperature and increasing Mn
concentration. This is shown in Fig.\,\ref{Fig1}(c) for the 10\,\% Mn-edge $(\frac{1}{4}+\Delta
q,\frac{1}{4}+\Delta q,0)$ scan, and in Fig.\,\ref{Fig1}(d,e) for the Mn and Ru-edge integrated peak
intensities (at 641 and 2968\,eV, respectively).

Altogether, the results in Fig.\,\ref{Fig1} demonstrate that the metal-insulator transition is accompanied by
a two-dimensional order with wavelength $2\sqrt{2}\,a_0$ and modulation vector $\boldsymbol{q}$ along the
diagonal of the $I$4$/mmm$ zone. The order is predominantly electronic and not structural (we did not observe
corresponding superlattice reflections in non-resonant low-temperature x-ray diffraction), nor associated
with the spatial ordering of the Mn impurities as evidenced by the doping independence of $\boldsymbol{q}$.
The nature of this electronic order, i.e. spin/charge/orbital, can be further clarified by the detailed
azimuthal-angle dependence of the integrated intensity of the Ru and Mn-edge $(\frac{1}{4},\frac{1}{4},0)$
peaks. The results for the 10\,\% Mn system are presented in Fig.\,\ref{Fig2}, together with a theoretical
angle dependence calculated for pure spin order \cite{Hill}, with spins parallel to the $c$-axis. The
agreement between measured and calculated azimuthal dependence implies the primarily magnetic nature of the
ordering, with average spin direction along the $c$-axis at both Ru and Mn sites.

To specify the exact pattern of the spin order, we should note that in addition to the above mentioned
forbidden superlattice peaks, also a $\boldsymbol{q}\!=\!(\frac{1}{4},\frac{5}{4},0)$ reflection was observed
in our neutron scattering study on the $x\!=\!0.05$ material \cite{unpublished}. Instead, no
$\boldsymbol{q}\!=\!(\frac{1}{2},\frac{1}{2},0)$ reflection was detected with either neutron or REXS at the
Ru edge (at the Mn edge this $\boldsymbol{q}$ value cannot be reached). These results enforce a very strong
constraint on the possible spin texture but still allow for more than one compatible spin pattern. Among
those, the most significant are the two highest symmetry ones, which due to the presence of domains (more
below) would give rise to identical REXS patterns: a {\it checkerboard} antiferromagnetic ordering of square
blocks of four parallel spins (Fig.\,\ref{Fig2}, left inset); alternatively, an antiferromagnetic alternation
of ferromagnetic {\it zigzag stripes} aligned diagonally with respect to the tetragonal $I$4$/mmm$ zone
(Fig.\,\ref{Fig2}, right inset). While the checkerboard pattern is isotropic with respect to the
crystallographic $a^*$ and $b^*$ axes of the {\it Bbcb} orthorhombic zone, the zigzag stripe pattern is not.
Since this electronic anisotropy is appealing in relation to the reports of a nematic fluid phase in
Sr$_3$Ru$_2$O$_7$ with $a^*\!-b^*$ resistive anisotropy in an applied magnetic field \cite{borzi,raghu}, let
us continue by focusing on the zigzag stripe order; note, however, that the following discussion on the
nature of the metal-insulator transition is independent of this choice.

\begin{figure}[t!]
\centerline{\epsfig{figure=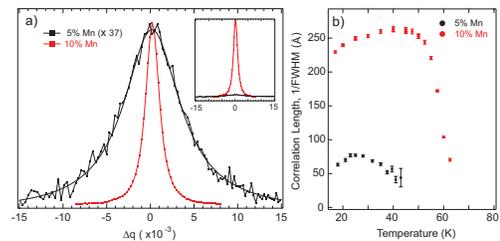,width=.75\linewidth,clip=}} \vspace{-0.25cm}\caption{(color
online). (a) Normalized Mn-edge $(\frac{1}{4}\!+\!\Delta q,\frac{1}{4}\!+\!\Delta q,0)$ momentum scans at
20\,K for 5\,\% (with Lorentzian fit) and 10\,\% Mn concentration; raw data are shown in the inset. (b)
Inverse FWHM of the momentum scans vs. temperature.}\label{Fig3}
\end{figure}

Because of their coincidence in temperature, it seems natural to associate the metal-insulator transition
with the onset of magnetic order induced by the Mn impurities. To further our understanding of the mechanism
of the metal-insulator transition, we have studied the doping and temperature dependence of the width and
intensity of the Mn-edge momentum scans, which provide an estimate of the magnetic-order correlation length
and of the fraction of participating Mn impurities. Fig.\,\ref{Fig3}(a) presents a comparison of the Mn-edge
$(\frac{1}{4}+\Delta q,\frac{1}{4}+\Delta q,0)$ momentum scans at 20\,K for $x\!=\!0.05$ and 0.1. As
evidenced by the combination of normalized (main panel) and raw data (inset), both the integrated intensity
and peak width are strongly enhanced (by factors of 10 and 4, respectively) upon increasing $x$ from 5 to
10\,\%. The detailed temperature dependence of the Mn-edge correlation length, defined as the inverse of the
full-width half maximum (FWHM) of the momentum scans, is shown in Fig.\,\ref{Fig3}(b) for both samples
(similar correlation length results were obtained at the Ru-edge).

To start modelling this behavior, we should note that only the magnetically correlated Mn impurities
contribute to the $(\frac{1}{4},\frac{1}{4},0)$ scattering intensity at the Mn $L$-edge; calculations
assuming a random distribution of Mn sites show that if 100\,\% of the Mn impurities took part in the
long-range spin order, as Mn doping is increased from 5 to 10\,\% the diffraction peak integrated intensity
should increase by a factor of about 4, i.e. nearly quadratically in the number of scattering centers
\cite{maurits_intensity}. The observed 10-fold increase implies that a much smaller fraction of the Mn
moments are correlated for $x\!=\!0.05$ and suggests the following scenario for the emergence of long-range
magnetic order. At a sufficiently low temperature, each Mn impurity surrounds itself by a two-dimensional
spin-ordered island, within the metallic RuO$_2$ plane. When these islands begin to overlap, the RuO-mediated
exchange interaction can energetically favor a coherent spin arrangement between the islands and, as a
result, the Mn impurities will interfere constructively. In the dilute 5\,\% Mn regime, there are isolated Mn
impurities whose spins are not participating in the order and will not contribute to the superlattice
diffraction peak intensity; at 10\,\% Mn concentration, however, there are statistically hardly any isolated
Mn atoms and hence almost all them will contribute.
As experimentally observed, the diffraction peak intensity would increase much faster than quadratically with
increasing $x$.

\begin{figure}[t]
\centerline{\epsfig{figure=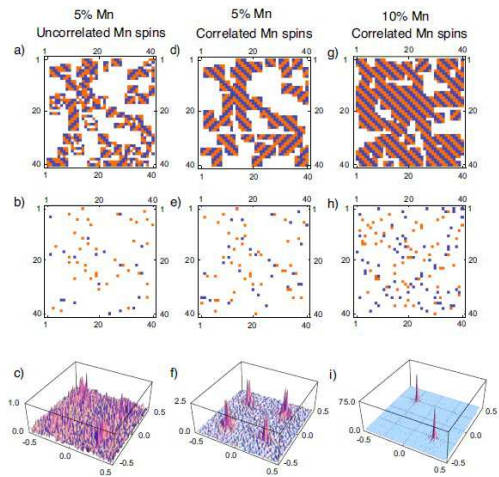,width=.76\linewidth,clip=}} \vspace{-0.25cm}\caption{(color
online). (a,d,g) Spin-ordered islands in the RuO$_2$ plane for various Mn contents and spin
correlations:\,(a) 5\,\% Mn, uncorrelated Mn spins; (d) 5\,\% Mn, correlated Mn spins; (g) 10\,\% Mn,
correlated Mn spins. (b,e,h) Corresponding location of Mn impurities with up/down (blue/red) spins. (c,f,i)
Reciprocal space map of the Mn scattering intensity generated by averaging over 200 random Mn-impurity
distributions, of the kind in (b,e,h), to reduce noise.}\label{SimFig}
\end{figure}

It might appear surprising that a REXS signal is obtained from dilute, randomly distributed Mn impurities. One should realize, however, that
constructive interference is dependent not on the spatial location of the impurities, provided that the Mn atoms occupy substitutional Ru sites,
but rather on the proper phase relation between their magnetic moments. This can be demonstrated with a simulation based on the correlation
lengths obtained from the inverse of the FWHM of the experimental momentum scans ($\sim$55 and 235\,\AA \ for 5 and 10\,\% Mn concentration,
respectively, at 20K). Fig.\,\ref{SimFig}(a) presents a 40$\times$40 RuO$_2$ lattice in real space, with 5\,\% Mn sites each inducing a
4$\times$4 unit of the zigzag stripe pattern (the simulation would identically apply to the case of the checkerboard ordering). Here, no
correlation between the Mn spins has been imposed. As in Fig.\,\ref{Fig2}, blue and red squares in Fig.\,\ref{SimFig}(a) represent up and down
spins along the $c$-axis, while the white patches are regions of the RuO$_2$ plane where no magnetism has been induced [the position of the Mn
impurities and their corresponding spins are shown in Fig.\,\ref{SimFig}(b)]. Due to the lack of correlation between the Mn spins, the  Mn-edge
$(\pm\frac{1}{4},\pm\frac{1}{4},0)$ peaks are extremely broad and weak, buried in the noisy background, as shown by the reciprocal space map of
the scattering intensity in Fig.\,\ref{SimFig}(c). The situation is very different for correlated Mn spins, as shown in
Fig.\,\ref{SimFig}(d,e,f) which presents the same sequence as panels (a,b,c), still for $x\!=\!5$\,\%, but for an average cluster size of
55\,\AA \ ($\sim$13 lattice spacing) as observed experimentally. The superlattice $(\pm\frac{1}{4},\pm\frac{1}{4},0)$ peaks are clearly
discernible above the noise in Fig.\,\ref{SimFig}(f). Note that this particular spin pattern also gives rise to
$(\frac{2n+1}{4},\frac{2n+1}{4},0)$ diffraction peaks, with $n$ being an integer, but for the present discussion it is sufficient to restrict
the field of view to $q_x, q_y\!=\!\pm 0.5\,\pi/a_0$. For $x\!=\!10$\,\% and $\sim$235\,\AA \ magnetic islands ($\sim$57 lattice spacing), we
can observe in Fig.\,\ref{SimFig}(g,h,i) sharp diffraction spots with a peak height 30 times that of $x\!=\!5$\,\%, in good agreement with the
experimentally observed $\sim$37-fold increase (Fig.\,\ref{Fig3}a). We should also note that $(\frac{1}{4},-\frac{1}{4},0)$ and
$(-\frac{1}{4},\frac{1}{4},0)$ peaks are absent because the domain size is now larger than the lattice cell used for the simulation; in the real
experiments they would be still visible.

We can now summarize our findings and try to establish a connection between long-range spin order and
metal-insulator transition in Sr$_3$(Ru$_{1-x}$Mn$_x$)$_2$O$_7$, and possibly the magnetic fluctuations and
nematic fluid behavior of the parent compound. Since Sr$_3$Ru$_2$O$_7$ does not show any long-range magnetic
order, it is clear that the latter is induced by the $S\!=\!2$, 3$d$-Mn$^{3+}$ impurities \cite{mathieu,
hossain_PRL}.
Nevertheless, the ordering is independent of the precise 5-10\,\% Mn concentration, suggesting that the role
of Mn is primarily that of triggering and/or stabilizing an instability incipient in the parent compound.
Indeed, strong two-dimensional spin fluctuations have been observed in Sr$_3$Ru$_2$O$_7$ in zero field, which
appear to cross over from ferro to antiferromagnetic upon reducing the temperature below 20\,K
\cite{capogna}; and also a momentum-dependent spin anisotropy induced by spin-orbit coupling should be
expected, as in Sr$_2$RuO$_4$ \cite{maurits}. For both checkerboard and zigzag stripe spin patterns, the
insulating behavior would result from the interaction of the propagating carriers with the ordered spin
background. Note, however, that the zigzag stripe order would exhibit a marked transport anisotropy even in
the absence of any applied field, with high and low conductivity behavior, respectively, along and
perpendicular to the ferromagnetic stripes. In this latter case the macroscopic metal-insulator transition
would thus depend on the formation of magnetic domains with different orientation. One might speculate that
the transport anisotropy expected for a single zigzag stripe domain in Sr$_3$(Ru$_{1-x}$Mn$_x$)$_2$O$_7$, and
the one detected in the parent compound in magnetic fields \cite{borzi}, could be fingerprints of the same
nematic fluid instability \cite{raghu}; this however will require further scrutiny.

It should be emphasized that Sr$_3$(Ru$_{1-x}$Mn$_x$)$_2$O$_7$ in the dilute Mn substitution regime provides
a unique opportunity to study with great accuracy the role of impurities and disorder in inducing phase
separation, percolative, and glassy behavior in the general class of correlated oxides. For instance, it is
interesting to note that the effective percolation threshold is achieved at $\lesssim\!5$\,\% Mn
concentration \cite{mathieu}; in terms of magnetically correlated sites, as shown in Fig.\,\ref{SimFig}(a,d)
this corresponds to 50.2\,\% of the RuO$_2$ plane, thus remarkably close to the threshold for classical
metallic percolation on a square lattice which is suppressed at a 50\,\% mixture of metallic and insulating
bonds \cite{Kirkpatrick}. Furthermore, the non-monotonic temperature dependence of the correlation length
presented in Fig.\,\ref{Fig3}(b) has also been seen in reentrant spin glasses, due to the interplay of
competing interactions and disorder \cite{Maletta,Aeppli}. Finally, speculating on the much higher resonant
enhancement observed at the Mn as compared to the Ru absorption edge (an at-least 36-fold enhancement should
be expected due to the cross-section difference between 2$p$-3$d$ and 2$p$-4$d$ excitations), one might
imagine using random impurities as a highly sensitive probe of spin/charge/orbital order of a host system
that is difficult to access via the majority lattice population.

We thank M.Z. Hasan for the use of the ALS scattering chamber. This work is supported by ALS (M.A.H), Sloan
Foundation (A.D.), CRC Program (A.D., G.A.S.), NSERC, CFI, CIFAR, and BCSI. ALS is supported by the U.S. DOE
Contract No. DE-AC02-05CH11231.

\vspace{-0.5cm}

\bibliographystyle{plain}

\end{document}